\documentclass[letterpaper,12pt]{article}
\usepackage{amssymb}
\usepackage{amsfonts}
\usepackage{amsmath}
\usepackage{amsthm}
\usepackage{color}
\usepackage{url}

\newcommand{\dont}[1]{}

\def\R{\mathbb{R}}
\def\endproof{\hfill\diamondsuit}

\def\sF{{\mathcal F}}

\def\sL{{\mathcal L}}

\def\sD{{\mathcal D}}

\def\sC{{\mathcal C}}
\def\L{\mathbb{L}}

\def\F{\mathbb{F}}
\def\E{\mathbb{E}}

\def\sF{\mathcal{F}}
\def\P{\mathbb{P}}

\def\Q{\mathbb{Q}}

\def\sE{{\mathcal E}}
\def\N{\mathbb N}

\numberwithin{equation}{section}
\theoremstyle{plain}                
\newtheorem{theorem}{Theorem}[section]
\newtheorem{lemma}[theorem]{Lemma}

\newtheorem{corollary}[theorem]{Corollary}
\theoremstyle{definition}           

\newtheorem{example}[theorem]{Example}

\newtheorem{assumption}[theorem]{Assumption}
\theoremstyle{remark}               

\begin{document}
\begin{center}
{\LARGE \bf Horizon dependence of utility optimizers in incomplete models}

\ \\ \ \\
Kasper Larsen\\ Department of Mathematical Sciences, \\
  Carnegie Mellon University,\\ Pittsburgh, PA 15213 \\ {\tt
    kasperl@andrew.cmu.edu}

\ \\ 
Hang Yu\\ Department of Mathematical Sciences, \\
  Carnegie Mellon University,\\ Pittsburgh, PA 15213 \\ {\tt
    hangy@andrew.cmu.edu}

\ \\\vspace{+5pt} \today \vspace{+5pt}
\end{center}
\begin{abstract}
This paper studies the utility maximization problem with changing time horizons in the incomplete Brownian setting. We first show that the primal value function and the optimal terminal wealth are continuous with respect to the time horizon $T$. Secondly, we exemplify that the expected utility stemming from applying the $T$-horizon optimizer on a shorter time horizon $S$, $S < T$, may not converge as $S\uparrow T$ to the $T$-horizon value. Finally, we provide necessary and sufficient conditions preventing the existence of this phenomenon.
\end{abstract}

\vspace{+20pt}


\newpage

\section{Introduction and summary}

We consider an investor maximizing expected utility of terminal wealth in a general incomplete Brownian based framework. We are interested in the stability part of Hadamard's well-posedness requirements and we will consider two continuity questions related to the investment horizon parameter $T\in[0,\infty)$. Our goal is to identify models which are stable in the sense that the following questions can be answered negatively. 

We first pose the question
\begin{center}
\emph{Does a (marginal) misspecification of the investment horizon significantly influence the investor's optimal strategy?}
\end{center}
We let $X_t^{(T)}$ denote the optimal wealth process at time $t\in[0,T]$ and we let $U$ be the investor's utility function. Our mathematically interpretation of the above is question is whether or not we have
\begin{align}\label{q1}
\lim_{K\to T}\;\E[U(X_K^{(K)})] = \E[U(X_T^{(T)})]?
\end{align}
Our first main result provides an affirmative anwer in the general incomplete Brownian setting without imposing any additional conditions on the model. 

The second question we seek to answer is
\begin{center}
\emph{Can the optimal $T$-horizon strategy be (marginally) terminated pre-maturely without the investor incurring a loss?}
\end{center}
The mathematical quantification of this question is whether or not we have
\begin{align}\label{q2}
\lim_{K\uparrow T}\;\E[U(X_K^{(T)})] = \E[U(X_T^{(T)})]?
\end{align}
Here $X^{(T)}$ denotes the $T$-horizon optimal wealth process and $X^{(T)}_K$ denotes its value at time $K$, $K<T$. We explicitly exemplify that \eqref{q2} does not hold in general. Specifically, for the negative power investor $U(a):= a^p/p$, $p<0$ and $a>0$, we construct a complete financial model where \eqref{q2} fails. We then develop sufficient conditions on the utility function alone as well as combined sufficient conditions on the utility function and the market structure for \eqref{q2} to hold in the incomplete Brownian setting. 

Alternatively we could consider models which explicitly incorporate uncertainty about the investment horizon. Such problems fall under the theory of robust utility maximization and typically involve a min-max objective, see e.g., the textbook \cite{FS02} and the references therein. These robust models seek to provide a strategy which works well for a variety of parameters. This is in contrast to our setting where the investor firmly believes in a fixed time horizon and we are interested in various continuity properties of the corresponding optimizer seen as a function of the horizon parameter.

In continuous time and state settings the problem of maximizing expected utility of terminal wealth dates back to Merton's original works. The general existence of optimizers in the complete Brownian-based setting is provided by the martingale method developed in \cite{CH89} and \cite{KLS87}. \cite{KLSX91} extend this method to the incomplete Brownian setting by introducing duality. Finally, \cite{KS99} and \cite{KS03} settle the question in the general semimartingale setting. 

The two questions raised above are about the sensitivity of optimizers with respect to the time horizon parameter $T$ and there exist several related research directions. Turnpike results show that if two utility functions for large wealth levels align then the corresponding optimizers converge as $T\to \infty$, see e.g., \cite{DRB99} and the references therein. \cite{FS08} and \cite{GR08} use ideas from the theory of large deviations to characterize the long-run optimizer ($T\to \infty)$ in various specific (Markovian) settings. Finally, we mention \cite{CS09} who establish the semimartingale structure of the collection of optimal terminal wealths $(X^{(T)}_T)_{T>0}$ in a general framework.

Our answer to the first question raised above partly rests on a variant of the duality results developed in \cite{LZ07}. The second question is complicated by the fact that $X_K^{(T)}$ is typically strictly suboptimal for non-myopic investors facing the investment horizon $K$ with $K<T$. As a consequence, we do not have a corresponding dual formulation which complicates the proofs (all proofs are in the Appendix).

We base our analysis on a probability space $(\Omega,\sF,\F,\P)$ where 
$\F:=(\sF_t)_{t\in[0,\overline{T}]}$ is the standard augmented filtration generated by a two dimensional Brownian motion $(B,W)$ with $\overline{T}\in[0,\infty)$ and we assume $\sF= \sF_{\overline{T}}$. 
$\sL^k$ denotes the set of measurable and adapted processes $\nu$ satisfying 
$$
\P\left(\int_0^{\overline{T}} |\nu_u|^k du <\infty\right)=1,\quad k=1,2.
$$ 
For a local martingale $M$, we denote by $\sE(M)_t$, $t\in[0,\overline{T}]$, the stochastic Dol\'eans-Dade exponential of $M$ with $\sE(M)_0=1$. Finally, for $a\in\R$ we define $a^-:= \max(-a,0)$ and $a^+:= \max(a,0)$. 

\section{The modeling framework}

\subsection{The financial market}
The market consists of the money market account $S_t^{(0)}$ with dynamics
\begin{align}\label{S0}
dS_t^{(0)} :=S_t^{(0)}r_t dt,\quad t\in
(0,\overline{T}) \quad S_0^{(0)} :=1,
\end{align}
for a nonnegative interest rate process $r\in\sL^1$. The single risky security has dynamics
\begin{align}\label{stock}
dS_t := S_t\Big(\mu_t dt + \sigma_t dB_t\Big),\quad t\in
(0,\overline{T}),\quad S_0 := 1.
\end{align}
We refer to $\mu\in\sL^1$ as the drift process and we call $0<\sigma\in\sL^2$ the volatility process. It is straightforward to extend the following results to allow for a $d$-dimensional Brownian motion driving $n$ risky stocks.  We refer to the ratio $\lambda:= (\mu-r)/\sigma$ as the market price of risk process. It is important to note that since the pair $(r,\lambda)$ is allowed to depend on the $W$-Brownian motion, the market $(S^{(0)},S)$ is in general incomplete and does not possess any Markovian structure.

The minimal martingale density $Z^\lambda$ defined by the Dol\'{e}ans-Dade
exponential 
\begin{align}\label{mmgd}
Z^\lambda_t := \sE(-\lambda \cdot B)_t:=
\exp\left(-\int_0^t \lambda_udB_u -\frac12\int_0^t
\lambda^2_udu\right),\quad t\in [0,\overline{T}],
\end{align}
is a strictly positive supermartingale for $\lambda\in\sL^2$. We refer to \cite{FS10} for more information about the minimal  density $Z^\lambda$.

In order to ensure that \eqref{S0}, \eqref{stock}, and \eqref{mmgd} are well-defined we assume throughout the paper that the following regularity condition is satisfied.

\begin{assumption}\label{ass1} The processes $(r,\mu,\sigma)$ satisfy the following: $\sigma\in\sL^2$ is strictly positive, $r\in\sL^1$ is nonnegative, and $\lambda := (\mu-r)/\sigma\in\sL^2$ is such that $Z^\lambda$ is a genuine martingale. 

$\endproof$

\end{assumption}

In order to be in the familiar setting of no free lunch with vanishing risk (see \cite{DS94}) we assume that $Z^\lambda$ is a genuine martingale\footnote{In the case $r_t:=0$, i.e., $S^{(0)}_t =1$ for all $t\in[0,\overline{T}]$, \cite{DS98} illustrate that Assumption \ref{ass1} is strictly stronger than the no free lunch with vanishing risk condition.}. Cauchy-Schwartz's inequality ensures that $\mu \in \sL^1$ under Assumption \ref{ass1}, hence, the dynamics \eqref{stock} are well-defined. 

\subsection{The investor's problem}

The investor's wealth at time $t\in[0,\overline{T}]$ is denoted by $X_t = X^{(x,\pi)}_t$ where $x>0$ is the initial endowment and $\pi$ denotes a wealth fraction $\pi$ invested in the risky asset $S$. The self-financing wealth dynamics are given by
\begin{align}\label{def:wealth}
dX_t = X_t r_t dt + X_t\pi_t\Big( (\mu_t-r_t) dt + \sigma_t dB_t\Big),\quad t\in(0,\overline{T}).
\end{align}
To ensure that these dynamics are well-defined we require $\pi$ to satisfy that $\pi\sigma\in\sL^2$. In this case, the regularity imposed on $(r,\lambda)$ in Assumption  \ref{ass1} ensures together with Cauchy-Schwartz's inequality that we have $\pi(\mu-r)\in\sL^1$. Furthermore, under Assumption \ref{ass1} there are no arbitrage opportunities in this class of strategies, see e.g., \cite{KLSX91}. 

The investor's preferences are modeled by a utility function $U$ defined on the positive axis. We assume that $U$ is strictly increasing, strictly concave, and continuously differentiable as well as satisfies the Inada and the reasonable asymptotic elasticity conditions, i.e., 
$$
\lim_{a\downarrow 0} U'(a) = +\infty,\quad \lim_{a\to +\infty} U'(a) = 0,\quad \limsup_{a\to +\infty}\; \frac{aU'(a)}{U(a)} <1.
$$
The main example of such a utility function is given by $U(a) := a^p/p$ for $p\in(-\infty,1)$ and $a>0$ with $p:=0$ interpreted as the myopic $log$-investor.  We assume that the investor seeks to maximize expected utility of terminal wealth
\begin{align}\label{def:primal}
u^{(K)}(x) := \sup_{\pi\,:\, \pi\sigma \in \sL^2} \E[U(X_K)] = \E[U(X^{(K)}_K)],\quad K\in[0,\overline{T}].
\end{align}
Implicitly in this definition is the usual convention that if $\pi$ renders 
$$
\E[U^+(X_K)]=\E[U^-(X_K)]=+\infty,
$$ 
we define $\E[U(X_K)] := -\infty$. 
We have explicitly augmented the value function $u^{(K)}(x)$ as well as the optimal terminal wealth process $X_t^{(K)}$, $t\in[0,K]$, with the maturity index $K\in[0,\overline{T}]$. Since shorter horizon strategies are always admissible for a longer horizon and since $r$ is assumed nonnegative, it is clear that $u^{(K)}(x)$ is nondecreasing in $K$. In Appendix \ref{sec:duality} we show how to adjust the duality theory developed in \cite{KS99} and \cite{KS03} to produce the unique optimizer $X^{(K)}$ in the present case of nonnegative interest raets.


\section{Continuity of the optimal behavior}

We first provide the left-continuity part for the first question \eqref{q1}.

\begin{theorem}\label{thm:optimal_infty} Under Assumption \ref{ass1} we have for $T\in[0,\overline{T}]$
\begin{enumerate}
\item $u^{(T)}(x)<\infty$ for all $x>0$ implies that we have $u^{(K)}(x) \uparrow u^{(T)}(x)$ as $K\uparrow T$ for all $x>0$,
\item $u^{(T)}(x)=+\infty$ for all $x>0$ implies that we have $u^{(K)}(x) \uparrow +\infty$ as $K\uparrow T$ for all $x>0$. 
\end{enumerate}
\end{theorem}

We next exemplify that even though $u^{(K)}(x)$ is finite for all $K\in[0,T)$ and all $x>0$ we can have $u^{(T)}(x) = +\infty$ for all $x>0$ where $T<\infty$. The following example is a simplified version of the celebrated Kim and Omberg model developed in \cite{KO96}, see also \cite{Liu07} for various affine extensions. \cite{KK04} list a sequence of other standard models which can display a similar exploding phenomenon.

\begin{example}\label{ex:KO}  We consider the complete model specification $r_t:=0, \,\sigma_t:=1$, and the OU-drift $\mu$ (equivalently, the market price of risk process $\lambda$) 
\begin{align}\label{def:KO}
d\mu_t := \kappa(\theta - \mu_t)dt + \beta dB_t, \quad t>0 \quad \mu_0\in\R,
\end{align}
for strictly positive constants $(\kappa,\theta,\beta)$. $U$ denotes the positive power investor's utility function, i.e., $U(a) :=a^p/p$ for $a\ge0$ and $p\in(0,1)$.

\begin{lemma}\label{lem:KO} Let $\mu$ be defined by \eqref{def:KO} and let $p\in(0,1)$. Then $Z_t^\mu$ defined by \eqref{mmgd} is a genuine martingale for all $t\in[0,\infty)$. Furthermore, for $\kappa>0$ sufficiently small there exists $T\in(0,\infty)$ such that mapping $E$ defined by
\begin{align}\label{KO:E}
[0,T)\ni K \to E(K) := \E\Big[ Z_K^{p/(p-1)}\Big], 
\end{align}
is continuous and finite-valued, however, converges to $+\infty$ as $K\uparrow T$.
\end{lemma}

Because $\mu$ is driven solely by $B$, the model $(S^{(0)},S)$ is complete and the martingale method from \cite{KLS87} and \cite{CH89} produces the (candidate) optimal terminal wealth  
$X^{(K)}_K := (yZ^\mu_K)^{1/(p-1)}$. Here $y = y(x)>0$ is the Lagrange multiplier corresponding to the investor's budget constraint, i.e., $y$ satisfies
$$
x = \E[Z_K^\mu(yZ^\mu_K)^{1/(p-1)}] = y^{1/(p-1)}\E[(Z^\mu_K)^{p/(p-1)}],\quad K\in[0,T).
$$
As pointed out and exemplified in \cite{KK04} care must given to verify that this candidate $X^{(K)}$ indeed is optimal (as stated on p.152 in \cite{KO96}, \cite{KO96} does not verify their HJB-argument). 

\begin{lemma}\label{lem:KO1} Let $\mu$ be defined by \eqref{def:KO}, let $p\in(0,1)$, and let $\kappa>0$ be small enough so Lemma \ref{lem:KO} can be applied. The primal value function is given by
\begin{align}
u^{(K)}(x) &= E(K)^{(1-p)}x^p/p,\quad x\ge0,\quad K\in[0,T),\label{KO:primal}
\end{align} 
where $E(\cdot)$ is defined by \eqref{KO:E}.
\end{lemma}
From \eqref{KO:primal} we indeed see that $u^{(K)}(x)$ is finite-valued for $K\in[0,T)$, however, explodes as $K\uparrow T$ by Lemma \ref{lem:KO} whenever $\kappa>0$ is sufficiently small. Finally, we mention that $\kappa:=0$ also displays this exploding phenomenon in finite time (as long as $p\in(0,1)$ and $\beta\neq0$).

$\endproof$

\end{example}

The next result completely settles the first question \eqref{q1} and constitutes our first main result.

\begin{theorem}\label{thm:optimal} Under Assumption \ref{ass1} and $u^{(T+\epsilon)}(x) <\infty$ for some $\epsilon >0$ such that $T+\epsilon \le \overline{T}$ we have
\begin{enumerate}
\item $u^{(K)}(\cdot) \to u^{(T)}(\cdot)$ uniformly on compacts of $(0,\infty)$ as $K\to T$,
\item $X^{(K)}_K \to X^{(T)}_T$ in $\P$-probability as $K\to T$.
\end{enumerate}
\end{theorem}

\section{Continuity and exiting pre-maturely}

\subsection{Formulation and a necessary condition}\label{counter}

In this section we consider the performance of the optimal investment strategy corresponding to the $T$-horizon on a shorter horizon $K$ with $K<T\le \overline{T}$, see \eqref{q2}. More specifically, we let  $X^{(T)}_K$ be the optimal wealth process at time $K$ corresponding to the $T$-horizon and to measure its performance up to time $K$ we are interested in the following quantity 
$$
u^{(T)}(K,x) :=  \E\left[U\left(X^{(T)}_K\right)\right],\quad X_0^{(T)}:=x,\quad K\in[0,T].
$$
This quantity measures the expected utility the investor obtains from following the $T$-horizon optimal strategy up to time $K$, $K<T$. From this definition it is clear that we have the following ranking
\begin{align}\label{ineq:ranking}
u^{(T)}(K,x) \le u^{(K)}(x) \le u^{(T)}(x),\quad K\in[0,T].
\end{align}

The previous section shows that $u^{(K)}(x)$ converges to $u^{(T)}(x)$ as $K\uparrow T$. A trivial application of Fatou's lemma and the path continuity of $(S^{(0)},S)$ show that when $U$ is uniformly bounded from below we also have $u^{(T)}(K,x) \to u^{(T)}(x)$ as $K\uparrow T$. In particular, this includes the positive power investor $U(a):= a^p/p$ for $p\in(0,1)$ and $a\ge0$, however, as the following result illustrates we do not in general have this convergence property.

\begin{theorem}\label{thm:neg} Let $U$ satisfy $U(0) = -\infty$, $U$ is uniformly bounded from above, and $U$ satisfies
\begin{align}\label{cond:neg_result}
\liminf_{a\to +\infty}\; aU'(a) =0.
\end{align}
Then there exists a complete financial model satisfying Assumption \ref{ass1} such that $u^{(T)}(x)$ is finite-valued for all $x>0$, however, $\lim_{K\uparrow T} u^{(T)}(K,x)=-\infty$ for some $x>0$.
\end{theorem}

This result shows that exiting an optimal strategy pre-maturely can have severe costs (measured in terms of loss in expected utility) for the investor.
We stress that this result covers the widely used class of negative power utility, i.e., $U(a) := a^p/p$ for $p<0$ and $a>0$. The assumption of $U$ being bounded from above only serves to ensure finiteness of the primal value function $u^{(T)}(x)$. 

Finally, we will try to give some interpretation of the construction underlying Theorem \ref{thm:neg}. First of all, since Assumption \ref{ass1} holds, there is no arbitrage in the underlying model, however, the model construction is inspired by the classical doubling strategy. The idea is that some ``good event'' is bound to happen before time $T$ but its timing can be arbitrarily close to $T$. Therefore, the $T$-horizon investor eventually receives this ``good payoff'', however, during $[0,T)$ the strategy leads to wealth levels very close to zero. Consequently, when an investor with $U(0)=-\infty$ is forced to liquidate the position prematurely, the resulting expected utility can be arbitrary negative. On the other hand, if the strategy is held to maturity $T$ the ``good event'' is realized bringing the expectation up to a finite value.

\subsection{Sufficient conditions}

Our first result places sufficient conditions on the utility function alone.

\begin{theorem} \label{thm:suf_U} Under Assumption \ref{ass1}, $u^{(T)}(x)<\infty$ for $x>0$, $U(0) = -\infty$,
\begin{align}\label{cond:suf_U}
\inf_{a>0}\; aU'(a) >0,\quad \text{and}\quad  \limsup_{a\downarrow 0} \;aU'(a) <+\infty,
\end{align}
we have $u^{(T)}(K,x) \to u^{(T)}(x)$ as $K\uparrow T$ for $x>0$.
\end{theorem}
This result includes the myopic $log$-investor where $U(a):= \log(a)$, $a>0$. As usual we denote by $I(\cdot)$ the inverse function of $U'(\cdot)$. Condition  \eqref{cond:suf_U} can then equivalently be stated as 
\begin{align}\label{cond:suf_UU}
\inf_{b>0}\; bI(b) >0,\quad \text{and} \quad \limsup_{b\to+\infty} \;bI(b) <+\infty.
\end{align}


The next results place conditions on both the utility function and the underlying market. The first result is stated for the negative power investor.

\begin{theorem}\label{thm:main} Under Assumption \ref{ass1} and $U(a) := a^p/p$,  $p<0$ and $a>0$: If there exist constants $\epsilon >0$ and $\gamma < p(1-p) <0$ such that
\begin{align}\label{cond:main}
\sup_{t\in[T-\epsilon,T]} \E\left[  \exp\left(-\gamma \int_t^T r_udu\right)\left(\frac{Z^\lambda_T}{Z^\lambda_t}\right)^\gamma\right]<\infty,
\end{align}
we have $u^{(T)}(K,x) \to u^{(T)}(x)$ as $K\uparrow T$ for $x>0$.
\end{theorem}
 
The proof of this result only hinges on the behavior of the utility function $U$ near zero and as a consequence we have the following corollary.

\begin{corollary}\label{cor:main} Under Assumption \ref{ass1}: Assume the utility function $U$ satisfies $u^{(T)}(x)<\infty$ as well as for some $p<0$
we have 
\begin{align}\label{cond:main2}
0 < \liminf_{a\downarrow 0}\; \frac{U'(a)}{a^{p-1}} \le \limsup_{a\downarrow 0}\;\frac{U'(a)}{a^{p-1}} <\infty.
\end{align}
Then condition \eqref{cond:main} ensures $u^{(T)}(K,x) \to u^{(T)}(x)$ as $K\uparrow T$ for $x>0$.
\end{corollary}

We next provide an easily verifiable Novikov-type of condition on the interest rate and market price of risk process which ensures that \eqref{cond:main} holds.

\begin{lemma}\label{lem:novikov} Assume there exist constants $\epsilon>0$ and $\delta >0$ such that $r\in\sL^1$ and $\lambda\in\sL^2$ satisfy that the function
$$
[T-\epsilon,T] \ni t \to \E\Big[\exp\big(\delta(r_t+ \lambda^2_t)\big)\Big] \in(0,\infty),
$$
is continuous. Then \eqref{cond:main} holds for any $\gamma<0$.
\end{lemma}
We conclude this section by presenting a simple example illustrating this lemma's usefulness.

\begin{example} Let $r_t:=0$ and let 
$v$ denote the Feller process with $v_0>0$ and the dynamics
\begin{align}\label{ex:feller}
dv_t := \kappa(\theta-v_t )dt + \beta\sqrt{v_t} \Big( \rho dB_t + \sqrt{1-\rho^2} dW_t\Big), \quad t>0, \quad 
\end{align}
for positive constants $(\kappa,\theta,\beta)$ and $\rho\in[-1,1]$. We model the market price of risk process as
\begin{align}\label{ex:mpr}
\lambda_t := \frac{C_0}{\sqrt{C_1 + v_t} }+ C_2\sqrt{C_3 + v_t}, \quad t\in[0,T],
\end{align}
where $C_0,...,C_3$ are nonnegative constants. For $C_0:=C_3:=0$ this model is the Chacko-Viceira model (see \cite{CV05}) and \cite{Kra05} presents a closed form solution for the power investor. The  specification \eqref{ex:mpr} for $\lambda$ is inspired by the extended affine class of models developed in \cite{CFK07} which is embedded by setting $C_1:=C_3:=0$. 

\cite{HK08} show that for $C_0\neq0$ the usual exponential affine structure considered in e.g., \cite{Liu07} of the involved Laplacian breaks down and consequently no known closed form solution exists for the power investor. However, the next observation ensures the validity of Lemma \ref{lem:novikov}.

\begin{lemma}\label{lem:ex} For $C_1>0$ the market price of risk process \eqref{ex:feller}-\eqref{ex:mpr} satisfies the condition of Lemma \ref{lem:novikov}.
\end{lemma}
$\endproof$
\end{example}

\section{Conclusion}

Based on Hardamard's classical well-posedness criteria, we have studied sensitivity of utility optimizers in the general incomplete Brownian setting. We have shown that the optimal terminal wealth depends continuously on the time horizon parameter without any further assumptions on the model. Subsequently, we considered the performance of the longer horizon optimizer on a shorter horizon and we illustrated that exiting an optimal strategy pre-maturely can have severe costs in terms of loss in expected utility. Finally, we provided easily verifiable conditions ruling out such behavior. 

As a side implication, we have extended the dual existence results of  \cite{KS99} to the cover the case of nonnegative stochastic  interest rates.

\appendix
\section{Proofs}

This Appendix contains all the proofs. Since the proof of Theorem \ref{thm:neg} is construction based we present it first. We then extend the duality theory of \cite{KS99} to include nonnegative stochastic interest rates before proceeding with the remaining proofs. The results related to stochastic exponentials being genuine martingales (Example \ref{ex:KO} and Lemma \ref{lem:ex}), are provided at the very end.

\subsection{Proof of Theorem \ref{thm:neg}}

We let $(t_n)_{n\in\N}\subseteq (0,1)$ be an increasing sequence of numbers converging to 1 and define the disjoint partition $(A_n)_{n\in\N}$ of $\Omega$ (up to a $\P$-null set) as follows:
\begin{align*}
&A_1:= \{B_{t_1} <0\},\\
&A_2:= \{B_{t_1} \ge 0\} \; \cap \{B_{t_2} - B_{t_1} < 0\},\\
&A_3:= \{B_{t_1} \ge 0\} \; \cap \{B_{t_2} - B_{t_1} \ge 0\} \; \cap \{B_{t_3} - B_{t_2} < 0\},
\end{align*}
and so on. We note that by the independence of Brownian increments we have $\P(A_k) = 1/2^k$ for $k\in \N$. The sequence of  random variables $(Y_k)_{k\in\N}$ is defined by
\begin{align*}
Y_k :=
  \begin{cases}
 a_k & \text{if } \;\;(B_{t_{k+1}}- B_{t_k}) \le \alpha_k\\
 b_k & \text{if } \;\;(B_{t_{k+1}}- B_{t_k}) > \alpha_k.
  \end{cases}
\end{align*}
The constants are assumed to satisfy $a_k\in(0,1)$ and $b_k>1$ whereas $\alpha_n$ is chosen such that $\E[Y_k] =1$ for $k\in\N$. We note that $Y_k$ is a positive $\sF_{t_{k+1}}$-measurable random variable independent of $\sF_{t_k}$ for $k\in\N$. We can then define the strictly positive L\'evy martingale by
$$
\xi_t := \E\left[\sum_{k=1}^\infty Y_k1_{A_k}\Big|\sF_t\right],\quad t\in[0,T].
$$

In what follows $U$ denotes a utility function satisfying \eqref{cond:neg_result}.  The proof is finished by showing how we can use $\xi_t$ to construct a complete financial market such that
\begin{align}
\lim_{K\uparrow T} u^{(T)}(K,x) = -\infty,\quad u^{(T)}(x) \in \R,\quad x>0,
\end{align}
for a specific choice of the sequences $(a_k)_{k\in\N}$ and $(b_k)_{k\in\N}$.

By the martingale representation theorem for the Brownian motion $B$ we can find an adapted measurable process $\lambda \in \sL^2$ such that
$$
d\xi_t = -\xi_t \lambda_t dB_t,\quad t\in(0,1),\quad \xi_0 =1.
$$
By defining the drift process $\mu_t := \lambda_t$, the volatility process $\sigma_t :=1$, and the interest rate $r_t:=0$ we see that $\xi$ is the density process $Z^\lambda$ defined by \eqref{mmgd} and consequently Assumption \ref{ass1} is satisfied.

To construct the two sequences $(a_k)_{k\in\N}$ and $(b_k)_{k\in\N}$ we first observe 
\begin{align*}
\E[Y_k1_{A_k}|\sF_{t_n}] =
  \begin{cases}
 Y_k 1_{A_k} & \text{for } \;\;n>k\\
 1_{A_n} & \text{for } \;\;n=k\\
 1_{C_n}\frac1{2^{k-n}} & \text{for } \;\;n<k,
  \end{cases}
\end{align*}
where we have defined the $\sF_n$-measurable set $C_n$ by
$$
C_n:= (B_{t_1}\ge 0) \cap (B_{t_2}-B_{t_1} \ge 0) \cap ... \cap (B_{t_k}-B_{t_{k-1}} \ge 0) = \left(\cup_{k=1}^n A_k\right)^c.
$$
By Tonelli's theorem for conditional expectations we have the relation 
$$
\xi_{t_n} = \sum_{k=1}^\infty \E[Y_k1_{A_k}|\sF_{t_n}] = \sum_{k=1}^{n-1} Y_k1_{A_k} +1_{A_n} + 1_{C_n}.
$$

By using the martingale method for complete Brownian based models, see \cite{CH89} and \cite{KLS87}, we know that the optimal terminal wealth with time horizon $T:=1$ satisfies
$$
X^{(1)}_1 = I(y\xi_1)\quad \text{where} \quad I(b) := (U')^{-1}(b),\quad b>0.
$$
Here the Lagrange multiplier $y>0$ is given by the investor's budget restriction, i.e., $y$ satisfies $x = \E[\xi_1 I(y\xi_1)]$ where $x>0$ is the investor's initial wealth. We will focus on the initial wealth $x_0$ such that $y=1$. We know that $X_t^{(1)}\xi_t$ is a martingale on $t\in[0,1]$ and therefore
$$
X_{t_n}^{(1)} = \frac{\E[X_1^{(1)}\xi_1|\sF_{t_n}]}{\xi_{t_n}} = \frac{\E[I(\xi_1)\xi_1|\sF_{t_n}]}{\xi_{t_n}},\quad n\in\N.
$$
Similarly to the above calculations we can compute the nominator to be
\begin{align*}
\E[I(\xi_1)\xi_1|\sF_{t_n}] &= \sum_{k=1}^\infty \E[I(Y_k)Y_k1_{A_k}|\sF_{t_n}]\\
&= \sum_{k=1}^{n-1}I(Y_k)Y_k1_{A_k} + \E[I(Y_n)Y_n]1_{A_n} + \sum_{k=n+1}^{\infty}\E[I(Y_k)Y_k]1_{C_n}\frac1{2^{k-n}}.
\end{align*}
We can then compute $u^{(1)}(t_n,x_0)$ for $n\in\N$ to be
\begin{align*}
\E[U(X_{t_n}^{(1)})] &= \E\left[U\left(\frac{\E[I(\xi_1)\xi_1|\sF_{t_n}]}{\xi_{t_n}}\right)\right]\\&=
\sum_{k=1}^{n-1}\E\left[U\Big(I(Y_k)\Big)1_{A_k}\right] + \E\Big[U\Big(\E[I(Y_n)Y_n]\Big)1_{A_n}\Big]\\& + \E\left[U\Big(\sum_{k=n+1}^\infty\E[I(Y_k)Y_k]\frac1{2^{k-n}}\Big)1_{C_n}\right]
\\&=
\sum_{k=1}^{n-1}\E\left[U\Big(I(Y_k)\Big)\right] \P(A_k)+ U\Big(\E[I(Y_n)Y_n]\Big)\P(A_n)\\& + U\Big(\sum_{k=n+1}^\infty\E[I(Y_k)Y_k]\frac1{2^{k-n}}\Big)\P(C_n).
\end{align*}
On the other hand $u^{(1)}(x_0)$ is given by
$$
\E[U(X^{(1)}_1)] = \E\left[U\Big(I(\xi_1)\Big)\right] = \sum_{k=1}^\infty\E\left[U\Big(I(Y_k)\Big)\right] \P(A_k).
$$
The goal is therefore to construct $(a_k)_{k\in\N}$ and $(b_k)_{k\in\N}$ such that 
$$
\lim_{k\to\infty} \E[Y_kI(Y_k)] = 0\quad \text{and}\quad \lim_{k\to \infty}U\Big(\E[Y_kI(Y_k)]\Big)\P(A_k)= -\infty.
$$
Provided that this can be done, we would also have $\sum_{k=n+1}^\infty\E[I(Y_k)Y_k]\frac1{2^{k-n}}$ converges to zero. All in all this construction would produce the limit
$$
\lim_{n\to\infty} u^{(1)}(t_n,x_0)= -\infty,
$$
whereas $u^{(1)}(1,x_0) \in \R$ and thereby conclude the proof. Since $U$ is unbounded from below we can find a sequence $(x_k)_{k\in\N}$ converging to zero such that $U(x_k) \P(A_k) = U(x_k)/2^k \to -\infty$. We then define $(a_k)_{k\in\N}$ and $(b_k)_{k\in\N}$ such that
$a_k \downarrow 0$ and $a_kI(a_k) < x_k/2$ (here we use \eqref{cond:neg_result}) and $b_k\uparrow +\infty$ such that $I(b_k) < x_k/2$. To summarize, we define $(\alpha_n)_{n\in\N}$ such that 
$$
\E[Y_n] = a_np_n+ b_n(1-p_n)=1,\quad p_n:= \P(B_{t_n+1}-B_{t_n}\le \alpha_n), \quad n\in\N,
$$ 
subsequently we define the density $(\xi_t)_{t\in[0,1]}$ in terms of $(Y_k)_{k\in\N}$, and finally we define the initial wealth by $x_0 := \E[\xi_1I(\xi_1)]<\infty$. 
Then we have
$$
\E[Y_kI(Y_k)] = a_kI(a_k)p_k + b_kI(b_k)(1-p_k) \le x_k \downarrow 0 \text{ as }k\to \infty,
$$
since both $p_k$ and $b_k(1-p_k)$ are less than one. The second requirement follows from the construction of $(x_k)_{k\in\N}$ and the increasing property of $U$
\begin{align*}
U\big(\E[Y_kI(Y_k)]\big)\P(A_k) &= U\big(a_kI(a_k)p_k + b_kI(b_k)(1-p_k) \big)/2^k\\& \le U(x_k)/2^k \to -\infty \text{ as }k\to \infty.
\end{align*}
$\endproof$

\subsection{Duality}\label{sec:duality}

The dual-based existence result in \cite{KS99} is derived under the assumption  $r_t:=0$ whereas \cite{KLSX91} rely on a uniform boundedness condition on $r$ as well as a smaller class of utility functions (excluding the negative power investors). This section explains how to adjust the dual approach of \cite{KS99} to cover the case of nonnegative interest rates for general utility functions.

A strictly positive progressively measurable process $Y$, $Y_0=1$, is called a supermartingale deflator if $XY$ is a supermartingale for any admissible wealth process $X$. For $\nu\in\sL^2$ It\^o's lemma shows that
\begin{align}\label{Ynu}
dY^\nu_t := -Y^\nu_t \Big( r_t dt + \lambda_t dB_t + \nu_t dW_t),\quad Y^\nu_0:=1,
\end{align}
is a supermartingale deflator. Since $r$ is assumed nonnegative $Y^\nu$ is a local supermartigale and to see that $Y^\nu$ is a  genuine supermartingale we let $(\tau_n)_{n\in\N}$ be a reducing sequence of stopping times. Fatou's lemma shows
$$
Y^\nu_s = \liminf_{n\to \infty} Y^\nu_{s\land \tau_n} \ge\liminf_{n\to \infty} \E[Y^\nu_{t\land \tau_n}|\sF_s] \ge \E[Y^\nu_t|\sF_s],
$$
for $0\le s\le t\le \overline{T}$. This supermartingale property is the key ingredient in the duality approach developed in \cite{KS99}.

\begin{lemma}\label{lem:Ynu} Under Assumption \ref{ass1}: For any supermartingale deflator $Y$ there exists $\nu\in\sL^2$ such that $Y_t\le Y_t^\nu$, $t\in[0,\overline{T}]$, where $Y_t^\nu$ is defined by \eqref{Ynu}.
\end{lemma}
\proof We define the discounted price system
\begin{align}\label{tildeS}
\tilde{S}^{(0)}_t := 1, \quad \tilde{S}_t := S_t/S^{(0)}_t,\quad t\in[0,\overline{T}],
\end{align}
and we denote by $\tilde{X}$ the corresponding discounted wealth process. Since $Y$ is a supermartingale deflator we also have $S^{(0)}Y \tilde{X}$ is a supermartingale for any $\tilde{X}$. Proposition 3.2 in \cite{LZ07} produces the representation $S^{(0)}Y = D Z^\lambda \sE(-\nu \cdot W)$ where $Z^\lambda$ is the minimal density \eqref{mmgd}, $\nu\in\sL^2$, and $D$ is a 
predictable, nonincreasing process with $D_0=1$ and $D_T>0$. Therefore,
$$
Y_t = D_t Z^\lambda_t \sE(-\nu \cdot W)_t/S_t^{(0)} \le Z^\lambda_t \sE(-\nu \cdot W)_t/S_t^{(0)} = Y_t^\nu, \quad t\in[0,\overline{T}].
$$
$\endproof$

We define the two sets of $\sF_T$-measurable random variables
\begin{align}\label{def:CD}
\sC&:= \{g\in\L^0_+(\P): g\le X_T, \;X_0=1\}, \quad \sD:=  \{h\in\L^0_+(\P): g\le Y^\nu_T\},
\end{align}
where $X$ denotes some admissible wealth process and $Y_t^\nu$ is given by \eqref{Ynu}. These sets generalize \cite{KS99}, equations (3.1) and (3.2) on p.912, to the case of nonnegative interest rates. The supermartingale property of $Y^\nu$ produces $\E[Y^\nu_T]\le 1$, i.e., $\sD$ defined by \eqref{def:CD} is a bounded set of $\L^1_+(\P)$.

\begin{lemma}\label{lem:bipolar} Under Assumption \ref{ass1}: The sets $\sC$ and $\sD$ are solid, convex and closed in probability. Furthermore, the sets are in bipolar relation, i.e., $\sD^\circ = \sC$ and $\sC^\circ=\sD$.
\end{lemma}

\proof This result is basically Proposition 3.1 from \cite{KS99} and here we only mention the few adjustments needed. The solidity and convexity of $\sD$ are clear (for the latter we can use Lemma \ref{lem:Ynu}). $\sD$'s closeness in probability follows as in \cite{KS99}, Lemma 4.1,  since $r\ge0$ ensures that $Y^\nu$ is a supermartingale. Hence, the Bipolar Theorem for $\L^0_+(\P)$ shows that $\sD = \sD^{\circ\circ}$. The inclusion $\sC\subseteq \sD^\circ$ is an immediate consequence of $XY^\nu$'s supermartingale property. 

The set of equivalent martingale measure $\Q$ for the discounted market \eqref{tildeS} is nonempty (because the minimal density $Z^\lambda$ is assumed to be a genuine martingale). Furthermore, the Radon-Nikodym derivative on $\sF_T$ of such a measure $\Q$ can be written as 
$$
\frac{d\Q}{d\P} = \exp(\int_0^T r_u du) Y_T^\nu,
$$
for some process $\nu = \nu^\Q\in\sL^2$. For any $X\in \L^0_+(\P)$ we define $\tilde{X} := X/S^{(0)}_T$ so we have the identity $\E[Y^\nu_T X] = \E^\Q[\tilde{X}]$. Therefore, if $\E[Y^\nu_T X]\le 1$ for all $\nu\in\sL^2$ we can super-replicate $\tilde{X}$ in $(\tilde{S}^{(0)}, \tilde{S})$ by standard arguments, hence, we can super-replicate $X$ in $(S^{(0)}, S)$. This shows that $\sD^\circ\subseteq \sC$ implying $\sD^\circ = \sC$. Therefore, the set $\sC$ is solid, convex, and closed in probability. By taking polars we see $\sC^\circ = \sD^{\circ \circ} = \sD$ which  finishes the proof.

$\endproof$

Based on \cite{KLSX91}, the dual value function corresponding to the primal problem \eqref{def:primal} is defined by
\begin{align}\label{def:dual}
v^{(T)}(y) := \inf_{\nu \in \sL^2} \;\E[V(y Y_T^\nu)],\quad y>0,
\end{align}
where $V$ is the convex conjugate of the utility function $U$ defined by \begin{align}\label{def:V}
V(b) := \sup_{a>0} \Big( U(a) - ab\Big),\quad b>0. 
\end{align}
Since $U(\cdot)$ is nondecreasing and $V(\cdot)$ is nonincreasing it is clear that the primal value function \eqref{def:primal} and the dual value function \eqref{def:dual} can also be written as
$$
u^{(T)}(x) = \sup_{g\in\sC} \;\E[U(xg)],\quad v^{(T)}(y) = \inf_{h \in \sD} \;\E[V(y h)],
$$
for $x,y>0$. 
Furthermore, Lemma \ref{lem:Ynu} shows that including all supermartingale deflators $Y$ in the minimizing \eqref{def:dual} produces the same infimum. From \eqref{def:V} we see for any $\nu\in\sL^2$ and any terminal wealth $X_T$ we have
$$
V(yY^\nu_T) \ge U(X_T) + yX_TY^\nu_T,\quad \P\text{-almost surely}.
$$
From this inequality we get the standard weak-duality inequality
\begin{align}\label{ineq:weak_dual}
v^{(T)}(y) \ge \sup_{x>0} \Big( u^{(T)}(x) - xy\Big),\quad y>0.
\end{align}
We note that \eqref{ineq:weak_dual} holds irrespectively of whether or not the primal and dual value functions are finite-valued.  The following result extends the main result of \cite{KS99} to the case of nonnegative stochastic interest rates.

\begin{theorem}[Kramkov-Schachermayer]\label{thm:KS} Under Assumption \ref{ass1}: If $u^{(T)}(x)<\infty$ then for $y>0$ the dual minimizer $\nu^{(T)} = \nu^{(T)}(y)\in\sL^2$ exists, i.e., 
$$
v^{(T)}(y) = \E[V(y Y_T^{\nu^{(T)}})].
$$
The primal and dual value functions are continuously differentiable as well as mutual conjugates, i.e.,
\begin{align}
&v^{(T)}(y) = \sup_{x>0} \Big( u^{(T)}(x) - xy\Big),\quad y>0,\label{eq:strong_dual_v}\\& u^{(T)}(x) = \inf_{y>0} \Big( v^{(T)}(y) + xy\Big),\quad x>0.\label{eq:strong_dual_u}
\end{align}
The unique optimal terminal wealth $X^{(T)}$ exists and satisfies $U'(X^{(T)}_T) = yY_T^{\nu^{(T)}}$ where $y>0$ is the Lagrange multiplier corresponding to the investor's budget constraint, i.e., $\frac{\partial}{\partial x} u^{(T)}(x) =y$. Furthermore, $X^{(T)}Y^{\nu^{(T)}}$ is a uniformly integrable martingale.
\end{theorem}

\proof 
The result follows from the ``abstract version'' in Section 3 of \cite{KS99}. Properties (i) and (ii) in Proposition 3.1 in \cite{KS99} are ensured by Lemma \ref{lem:bipolar}. The last required property (iii) follows from $r\ge0$, hence, $1\in\sC$, whereas Markov's inequality and $\sD \neq \emptyset $ produce the boundedness in probability of $\sC$.

$\endproof$

\subsection{Remaining proofs}

We remark that no finiteness of the dual value function $v^{(K)}(y)$ for $K=T$ is assumed in the following lemma.

\begin{lemma} \label{lem:key_optimal} Under Assumption \ref{ass1}: If $v^{(K)}(y) <\infty$ for all $K\in[0,T)$, $T\le \overline{T}$, then we have
$$
\liminf _{K\uparrow T, \;y\to y_0}\; v^{(K)}(y) \ge v^{(T)}(y_0),\quad y_0>0.
$$
\end{lemma}

\proof By means of the closeness of $\sD$ established in Lemma \ref{lem:bipolar}, the proof only requires minor modifications to the proof of Lemma 3.7 in \cite{LZ07}.

$\endproof$

\begin{lemma}\label{lem2} Under Assumption \ref{ass1}: $u^{(T)}(x)=+\infty$ for all $x>0$ if and only if $v^{(T)}(y)=+\infty$ for all $y>0$.
\end{lemma}
\proof If $u^{(T)}(x)=+\infty$ the weak duality inequality \eqref{ineq:weak_dual} gives us $v^{(T)}(y)=\infty$ for $y>0$. On the other hand, if $u^{(T)}(x) <+\infty$ for some $x>0$ - equivalently for all $x>0$ by $U$'s concavity - we have by \eqref{eq:strong_dual_u} that there exists some $y_0>0$ such that $v^{(T)}(y_0)<\infty$.

$\endproof$

\proof[Proof of Theorem \ref{thm:optimal_infty}] For $x>0$ we have $u^{(K)}(x)$ is increasing in $K$ and as a consequence $v^{(K)}(y)$ is increasing in $K$ too. Therefore, Lemma \ref{lem:key_optimal} shows 
$$
\lim_{K\uparrow T, y\to y_0} v^{(K)}(y) = v^{(T)}(y_0).
$$
The proof of the first claim then follows by the conjugate relationship \eqref{eq:strong_dual_v}-\eqref{eq:strong_dual_u} between the primal value function and dual value function, see Proposition 3.9 in \cite{LZ07}. 

For the second claim we define the concave function
$$
\overline{u}(x) := \lim_{K\uparrow T} u^{(K)}(x),\quad x>0.
$$
If $\overline{u}(x)$ is finite for some - equivalently all $x>0$ - we get for $y>0$
\begin{align*}
v^{(K)}(y) = \sup_{x>0}\; \Big( u^{(K)}(x) - xy\Big)
 \le \sup_{x>0}\; \Big( \overline{u}(x) - xy\Big),
\end{align*}
which is finite for some $y_0>0$. This provides a uniform upper bound in $K\in[0,T)$. By Lemma \ref{lem:key_optimal} we therefore also have $v^{(T)}(y_0)<\infty$, however, this contradicts the conclusion of Lemma \ref{lem2}. 

$\endproof$

\proof[Proof of Theorem \ref{thm:optimal}] Since $u^{(K)}(x)$ is increasing $K$ we have that $u^{(K)}(x)$ is finite-valued implying that also $v^{(K)}(y)$ is finite-valued for all $K\in[0,T+\epsilon]$. We fix $y>0$ and define the function
$$
f(\nu,K) := \E[V(yY^\nu_K)],\quad \nu \in \sL^2, \quad K\in[0,T+\epsilon].
$$
By the above observation $f$'s effective domain is non-empty. Furthermore, since $Y^\nu_t$ is a nonnegative supermartingale and since $V$ is a convex  nonincreasing function we have for $0\le s\le t\le \overline{T}$ the relation
$$
V(yY^\nu_s) \le V\Big(\E[yY_t^\nu|\sF_s]\Big)\le \E[V(yY_t^\nu)|\sF_s],
$$
by Jensen's inequality. Since the right-hand-side is integrable for $\nu$ in $f$'s effective domain, $V(yY_t^\nu)$ is a continuous submartingale. By Theorem 1.3.13 in \cite{KS91} we therefore know that $K \to f(K,\nu)$ is right-continuous (and non-decreasing) on its effective domain. Consequently, we find the relation
$$
 \inf_{K>T}f(\nu,K) =\lim_{K\downarrow T}f(\nu,K) = f(\nu,T),
$$
on $f$'s effective domain. Since $v^{(K)}(y)$ is also nondecreasing in $K$ we have 
\begin{align*}
\lim_{K\downarrow T} v^{(K)}(y) &= \inf_{K> T} v^{(K)}(y) \\&= \inf_{K> T} \inf_{\nu\in\sL^2} f(\nu,K) \\&= \inf_{\nu\in\sL^2} \inf_{K> T}  f(\nu,K) \\&= \inf_{\nu\in\sL^2} f(\nu,T) = v^{(T)}(y). 
\end{align*}
This shows right-continuity of the dual value function which combined with Theorem \ref{thm:optimal_infty} (first part) gives us the continuity property of $v^{(K)}(y)$. Having established the  continuity of the dual value function $v^{(K)}(y)$ we can prove that the primal value function shares the same continuity property, see Proposition 3.9 in \cite{LZ07}. The procedure used in Lemma 3.6 in \cite{KS99} and Lemma 3.10 in \cite{LZ07} subsequently shows the continuity (in probability) of the optimal terminal wealths $X^{(K)}_K$, $K\in[0,T+\epsilon]$.

$\endproof$

\proof[Proof of Theorem \ref{thm:suf_U}] Since $u^{(T)}(x)<\infty$, Theorem \ref{thm:KS} produces the primal optimizer $X^{(T)}$, the dual optimizer $Y^{\nu^{(T)}}$, and their relation
$$
X^{(T)}_t = \frac{\E[Y^{\nu^{(T)}}_TI(yY^{\nu^{(T)}}_T)|\sF_t]}{Y^{\nu^{(T)}}_t},\quad t\in[0,T].
$$
In this expression $y>0$ denotes the Lagrange multiplier corresponding to the budget restriction related to the investor's initial wealth $x>0$, i.e., $y$ is implicitly given by $x = \E[Y^{\nu^{(T)}}_TI(yY^{\nu^{(T)}}_T)]$. From the first part of \eqref{cond:suf_U} - or equivalently the first part of \eqref{cond:suf_UU} - we can find $\epsilon >0$ such that $bI(b) \ge \epsilon$ for all $b>0$. This implies that we have
the lower bound
$$
X^{(T)}_t \ge \frac\epsilon{yY^{\nu^{(T)}}_t},\quad t\in[0,T].
$$
By the second part of \eqref{cond:suf_U} and  $U(0)=-\infty$ we can use the below Lemma \ref{lem1} to get
$$
\limsup_{a\downarrow0} \; \frac{U(a)}{\log(a)} <+\infty.
$$ This means that we can find $a_0\in(0,1)$ and $M<\infty$ such that $U(a)/\log(a) < M$ for all $a\le a_0$. As a consequence we have
\begin{align}\label{ineq:lowerbound}
U^-(X^{(T)}_t) \le U^-\left(\epsilon/{yY^{\nu^{(T)}}_t}\right) \le M\log^-\left(\epsilon/yY^{\nu^{(T)}}_t\right) + D,
\end{align}
where $D>0$ is some constant. To see that the right-hand-side of \eqref{ineq:lowerbound} is uniformly integrable we define the uniform integrability test function $\phi(a) := \epsilon e^a/y$ for $a\in\R$. Then we have
\begin{align*}
\E\left[\phi\left(\log^-\left(\epsilon/yY^{\nu^{(T)}}_t\right)\right)\right] &\le \E\left[\phi\left(-\log\left(\epsilon/yY^{\nu^{(T)}}_t\right)\right)\right] + \phi(0)\\
&= \E[Y^{\nu^{(T)}}_t] + 1, 
\end{align*}
which is uniformly bounded for $t\in[0,T]$. The uniform integrability then follows from the de la Vall\'ee-Poussin's criterion. This feature combined with Fatou's lemma applied to the positive parts $U^+(X^{(T)}_t)$ gives us
$$
\liminf_{t\uparrow T} u^{(T)}(t,x) = \liminf_{t\uparrow T} \E[U(X^{(T)}_t)] \ge \E[U(X^{(T)}_T)] = u^{(T)}(x),
$$
by path continuity of $(S^{(0)},S)$. The opposite inequality follows from \eqref{ineq:ranking}.

$\endproof$

\begin{lemma}\label{lem1} Let $f,g:(0,\infty)\to\R$ be two strictly increasing and continuously differentiable functions satisfying
$$
\lim_{a\downarrow 0} \;f(a) = \lim_{a\downarrow 0} \;g(a) = -\infty,\quad \limsup_{a \downarrow 0} \;\frac{f'(a)}{g'(a)} < +\infty.
$$
Then we have 
$$
\limsup_{a \downarrow 0}\; \frac{f(a)}{g(a)} < +\infty.
$$
\end{lemma}

\proof By assumption we can find $a_0\in(0,\infty)$ such that $f'(a)/g'(a)$ is uniformly bounded for $a\in(0,a_0)$. Therefore, $\big(f(a_0) - f(a)\big)/\big(g(a_0)-g(a)\big)$ is bounded for $a$ sufficiently small. The formula 
$$
\frac{f(a)}{g(a)} = \frac{f(a_0) - f(a)}{g(a_0)-g(a)} \frac{f(a)}{f(a)-f(a_0)} \frac{g(a)-g(a_0)}{g(a)},
$$
together with L'Hopital's rule applied to the last two terms on the right-hand-side produces the claim.

$\endproof$

\begin{lemma}\label{lem:supermg} Under Assumption \ref{ass1} and $u^{(T)}(x)<\infty$, $T\in[0,\overline{T}]$: The process
$$
\frac1{\sE(-\nu^{(T)} \cdot W)_t} \E\left[ Y^{\nu^{(T)}}_T I\Big(yY^{\nu^{(T)}}_T\Big)\Big|\sF_t\right],\quad t\in[0,T],
$$
is a supermartingale where $\nu^{(T)}$ denotes the optimal dual element and $y$ denotes the Lagrange multiplier corresponding to the investor's budget constraint.
\end{lemma}

\proof Since $u^{(T)}(x)<\infty$ the dual optimizer $\nu^{(T)}$ exists by Theorem \ref{thm:KS} and we have the martingale representation 
$$
X^{(T)}_t Z^\lambda_t \sE(-\nu^{(T)}\cdot W)_t/S^{(0)}_t = Y^{\nu^{(T)}}_tX^{(T)}_t = \E\left[ Y^{\nu^{(T)}}_T I\Big(yY^{\nu^{(T)}}_T\Big)\Big|\sF_t\right].
$$
It\^o's lemma ensures $X^{(T)}_t Z^\lambda_t/S^{(0)}_t$ is a local martingale which by nonnegativity is also a supermartingale. Dividing through produces the result. 

$\endproof$

\proof[Proof of Theorem \ref{thm:main}] Since $U$ is negative the dual minimizer $\nu^{(T)}$ exists by Theorem \ref{thm:KS}. The proof is finished by showing that \eqref{cond:main} ensures uniform integrability of the family $\{(X_t^{(T)})^p\}_{t\in[T-\epsilon,T]}$. The inverse of $U'$ is given by $I(b) = b^{1/(p-1)}$, $b>0$. Let $y>0$ be the Lagrange multiplier corresponding to the investor's budget constraint.  The optimal wealth process $X_t^{(T)}$ satisfies for $t\in[0,T]$
\begin{align*}
X_t^{(T)} &= \frac1{Y_t^{\nu^{(T)}}}\E\left[ Y^{\nu^{(T)}}_T I\Big(yY^{\nu^{(T)}}_T\Big)\Big|\sF_t\right] \ge \frac{S_t^{(0)}}{Z^\lambda_t}\E\left[ Z^\lambda_T I\Big(yY^{\nu^{(T)}}_T\Big)/S^{(0)}_T\Big|\sF_t\right]
\end{align*}
where the inequality follows by the supermartingale property proven in the previous Lemma \ref{lem:supermg}. We define the negative constant
$$
p' := \frac1{1/\gamma + 1/(p-1)} \in (p-1,p).
$$
By de la Vall\'ee-Poussin's criterion it therefore suffices to show that the family $\{(X_t^{(T)})^{p'}\}_{t\in[T-\epsilon,T]}$ is uniformly bounded in $\L^1(\P)$. Since $p'<0$ the above inequality gives us $\P$-a.s. for $t\in[T-\epsilon,T]$ the estimate
\begin{align*}
(X_t^{(T)})^{p'} &\le \frac{y^{p'/(p-1)}}{\left(Z^\lambda_t/S^{(0)}_t\right)^{p'}}\E\left[Z^\lambda_T/S^{(0)}_T\Big(Y^{\nu^{(T)}}_T\Big)^{1/(p-1)}\Big|\sF_t\right]^{p'}\\&\le \frac{y^{p'/(p-1)}}{\left(Z^\lambda_t/S^{(0)}_t\right)^{p'}}\E\left[\left(Z^\lambda_T/S^{(0)}_T\right)^{p'}\Big(Y^{\nu^{(T)}}_T\Big)^{p'/(p-1)}\Big|\sF_t\right]
\\&= y^{p'/(p-1)}\E\left[\left(\frac{Z^\lambda_TS_t^{(0)}}{Z^\lambda_tS^{(0)}_T}\right)^{p'} \Big(Y^{\nu^{(T)}}_T\Big)^{p'/(p-1)}\Big|\sF_t\right],
\end{align*}
where the second inequality follows from Jensen's inequality. We define conjugate exponents 
$$
q := \frac{p-1}{p'} >1,\quad q' := \frac{q}{q-1} = \frac{p-1}{p-1-p'} =\frac{\gamma}{p'},
$$
by the definition of $p'$. These exponents together with H\"older's inequality give us
\begin{align*}
\E[(X_t^{(T)})^{p'}] 
&\le y^{p'/(p-1)}\E\left[\left(\frac{Z^\lambda_TS_t^{(0)}}{Z^\lambda_tS^{(0)}_T}\right)^{p'} \Big(Y^{\nu^{(T)}}_T\Big)^{p'/(p-1)}\right]\\
&\le y^{p'/(p-1)}\E\left[\left(\frac{Z^\lambda_TS_t^{(0)}}{Z^\lambda_tS^{(0)}_T}\right)^\gamma\right]^{1/q'}\E\left[Y^{\nu^{(T)}}_T\right]^{1/q}\\
&\le y^{p'/(p-1)}\E\left[\left(\frac{Z^\lambda_TS_t^{(0)}}{Z^\lambda_tS^{(0)}_T}\right)^\gamma\right]^{1/q'}.
\end{align*}
The supermartingale property of $Y^{\nu^{(T)}}$ yields $\E[Y^{\nu^{(T)}}]\le 1$ ensuring the validity of the last inequality. The right-hand-side is uniformly bounded by \eqref{cond:main}.

$\endproof$

\proof[Proof of Corollary \ref{cor:main}] We use the same notation as in the previous proof. We will first show that 
\eqref{cond:main} ensures uniform integrability of the negative parts $\{U^-(X_t^{(T)})\}_{t\in[T-\epsilon,T]}$. Condition \eqref{cond:main2} ensures that we can find $\overline{x}>0$ and $0<\underline{M}\le \overline{M}<\infty$ such that
\begin{align}\label{M_bounds}
\underline{M} a^{p-1} \le U'(a) \le \overline{M}a^{p-1}\quad \text{for all}\quad a\in(0,\overline{x}].
\end{align}
By integrating we therefore see that $U(a)$ is bounded from below by an affine function of $a^p$ for small values of $a$. In particular, we can find positive constants $C_1$ and $C_2$ such that $U^-(a) \le C_1a^p + C_2$ for all $a>0$. By de la Vall«ee-Poussin's criterion the uniform integrability follows if we can find $\delta>0$ such that the family $\{U^-(X_t^{(T)})\}_{t\in[T-\epsilon,T]}$ is uniformly bounded in $\L^{1+\delta}(\P)$. We proceed as in the previous proof: The supermartingale property established in Lemma \ref{lem:supermg} and Jensen's inequality ($p<0$) give us for $t\in[T-\epsilon,T]$
\begin{align*}
\E[\big(U^-(X_t^{(T)})\big)^{1+\delta}] &\le C_3\E[(X_t^{(T)})^{p(1+\delta)}] + C_4\\
&\le C_3\E\left[\left(\frac{Z^\lambda_TS^{(0)}_t}{Z^\lambda_tS^{(0)}_T} I\Big(yY^{\nu^{(T)}}_T\Big)\right)^{p(1+\delta)}\right]+C_4,
\end{align*}
where $C_3,C_4$ are constants. The lower bound in \eqref{M_bounds} gives us
$$
I(b) \ge \left(\frac{b}{\underline{M}}\right)^{1/(p-1)} \quad \text{for all }\quad  b\ge \underline{M}\overline{x}^{p-1}.
$$
Since $p$ is negative we have the following upper bound $\P$-a.s. 
$$
\left(I\Big(yY^{\nu^{(T)}}_T\Big)\right)^{p(1+\delta)} \le \Big(I(\underline{M}\overline{x}^{p-1})\Big)^{p(1+\delta)} + \underline{M}^{p(1+\delta)/(1-p)}\Big(yY^{\nu^{(T)}}_T\Big)^{p(1+\delta)/(p-1)}.
$$
By combining these two estimates we see that $\E[(U^-(X_t^{(T)}))^{1+\delta}]$ is bounded from above by
\begin{align*}
C_5\E\left[\left(\frac{Z^\lambda_TS_t^{(0)}}{Z^\lambda_tS^{(0)}_T} \right)^{p(1+\delta)}\right]+C_6\E\left[\left(\frac{Z^\lambda_TS^{(0)}_t}{Z^\lambda_tS^{(0)}_T} \right)^{p(1+\delta)}\Big(Y^{\nu^{(T)}}_T\Big)^{p(1+\delta)/(p-1)}\right]+C_4,
\end{align*}
where $C_5,C_6$ are constants. The terms on the right-hand-side are uniformly bounded in $t$, $t\in[T-\epsilon,T]$,  by the same reasoning as in the previous proof. To conclude the proof we apply Fatou's Lemma on the positive parts to see
\begin{align*}
\liminf_{t\uparrow T}\;u^{(T)}(t,x) &= \liminf_{t\uparrow T}\;\Big(\E[U^+(X_t^{(T)})] - \E[U^{-}(X_t^{(T)})]\Big)\\
&\ge \E[U^+(X_T^{(T)})] - \E[U^{-}(X_T^{(T)})] = u^{(T)}(x),
\end{align*}
whereas the opposite inequality follows from \eqref{ineq:ranking}.

$\endproof$

\newpage
\proof[Proof of Lemma \ref{lem:novikov}] Let $t\in[T-m,T]$ for some $m>0$. We have the following chain of inequalities
\begin{align*}
\E\left[ \left(\frac{Z_T^\lambda S^{(0)}_t}{Z^\lambda_tS^{(0)}_T}\right)^\gamma\right] &= \E\left[\exp\left(-\gamma\int_t^T \lambda_udB_u -\frac12\gamma \int_t^T (\lambda^2_u+2r_u) du\right)\right] \\&\le \E\left[\exp\left(-2\gamma\int_t^T \lambda_udB_u -2\gamma^2 \int_t^T \lambda^2_u du\right)\right]^{1/2}\times\\&\quad\quad \quad  \E\left[\exp\left(\int_{T-m}^T\Big((2\gamma^2-\gamma) \lambda^2_u -2\gamma r_u\Big)du\right)\right]^{1/2}\\
&\le \E\left[\exp\left(\int_{T-m}^T\Big((2\gamma^2-\gamma) \lambda^2_u -2\gamma r_u\Big)du\right)\right]^{1/2}\\
&\le \left[\frac1m \int_{T-m}^T \E\left[\exp\left(m\Big((2\gamma^2-\gamma) \lambda^2_u-2\gamma r_u\Big)\right)\right]du\right]^{1/2}.
\end{align*}
The first inequality follows from Cauchy-Schwartz's inequality, the second follows by the supermartingale property of the stochastic exponential whereas Jensen's inequality gives us the final estimate. To finish the proof we choose $m$ so small that $m(2\gamma^2-\gamma)\le \delta$, $-m2\gamma\le \delta$, and $m\le \epsilon$ (recall $\gamma<0$). Then the result follows since a continuous function on a compact interval is uniformly bounded.

$\endproof$

\subsection{On Example \ref{ex:KO} and Lemma \ref{lem:ex}}

The following proofs are based on iterative technique presented in \cite{LS77}, Section 6.2.

\proof[Proof of Lemma \ref{lem:KO}] 

Since $\mu$ is an OU-process we have $\mu_t \sim \mathcal{N}\big(\text{mean}(t),\text{var}(t)\big)$ for mean and variance functions 
$$
\text{mean}(t):= e^{-\kappa t}\mu_0 + \theta(1-e^{-\kappa t}), \quad \text{var}(t):= \beta^2(1-e^{-2\kappa t})/2\kappa, \quad t\ge0.
$$
Let us first verify that the exponential local martingale $Z^\lambda = Z^\mu$ defined by \eqref{mmgd} indeed is a genuine martingale, see Assumption \ref{ass1}. We consider a finite partition $\Delta = \Delta_n := \overline{T}/n$ for some $n\in\N$ and some $\overline{T}>0$. By Tonelli's Theorem and Jensen's inequality we have
$$
\E\left[e^{\frac12 \int_{i\Delta}^{(i+1)\Delta} \mu_t^2dt}\right] \le \frac1\Delta\int_{i\Delta}^{(i+1)\Delta} \E\left[e^{\frac12 \Delta\mu_t^2}\right]dt.
$$
The inner expectation is finite whenever $n$ is so large that $v(t)< n/\overline{T}$. This is ensured by choosing $n\ge \overline{T}\beta^2/2\kappa$. Consequently, Novikov's condition is valid on each subinterval which combined with iterative expectations gives us
$$
\E[Z^\mu_{\overline{T}}] = \E\big[\E[Z^\mu_{\overline{T}}|\sF_{\overline{T}-\Delta}]\big] =\E[Z^\mu_{\overline{T}-\Delta}] = ... =1.
$$ 
Since this holds for any $\overline{T}>0$ the genuine martingale property of $Z^\mu$ on $[0,\infty)$ follows.

We then consider the coupled system of ODEs
\begin{align}
c'(s) &:= \frac{p}{(p-1)^2} -2 c(s)\left(\kappa+\frac{\beta p}{p-1}\right) +\beta^2c(s)^2,\quad   c(0):=0,\label{ode:c}\\
b'(s) &:= \kappa\theta c(s) +b(s)\left(\beta^2c(s) - \kappa -\frac{\beta p}{p-1}\right),\quad b(0):=0,\nonumber \\
a'(s) &:= \kappa\theta b(s) + \frac12 \beta^2\big(b(s)^2 +c(s)\big),\quad a(0):=0.\nonumber 
\end{align}
For $\kappa>0$ sufficiently small the Riccati equation \eqref{ode:c} has a well-defined (finite) solution up to some finite explosion time $T\in(0,\infty)$. Specifically, since $p\in(0,1)$, we can choose $\kappa>0$ so small that the discriminant
$$
4\left(\kappa+\frac{\beta p}{p-1}\right)^2 - 4\frac{\beta^2 p}{(p-1)^2},
$$
is negative. We then obtain the Tangent-solution presented in Appendix in \cite{KO96} which explodes continuously at some positive finite time $T\in(0,\infty)$. Consequently, the entire ODE-system for $(a,b,c)$ has well-defined finite solutions up to this explosion time $T$.
For $0\le t \le K<T$ we can then define the process
$$
M_t := Z_t^{p/(p-1)}\exp\Big(a(K-t) + b(K-t)\mu_t + 
c(K-t)\mu_t^2/2\Big).
$$
It\^o's lemma produces the following local martingale dynamics 
$$
dM_t = M_t \Big( b(K-t)\beta + \mu_t\big(c(K-t)\beta -p/(p-1)\big)\Big)dB_t.
$$
Since both $b(\cdot)$ and $c(\cdot)$ are uniformly bounded on $[0,K]$, $K<T$, the above localization argument also produces the genuine martingale property of $M_t$, $t\in[0,K]$. Consequently, we obtain via the initial conditions for $(a,b,c)$ the representation
$$
E(K) := \E\Big[ Z_K^{p/(p-1)}\Big]=\exp\Big(a(K) + b(K)\mu_0 + 
c(K)\mu_0^2/2\Big), \quad K<T.
$$
Since the Riccati equation \eqref{ode:c} explodes continuously to $+\infty$, as $K\uparrow T$, this representation shows that $\lim_{K\uparrow T} E(K) =+\infty$.

$\endproof$

\proof[Proof of Lemma \ref{lem:KO1}] The following duality argument is standard. For $U(a):=a^p/p$ we have $V(b) =\frac{1-p}{p} b^{p/(p-1)}$. Let $X_K$ be the terminal value of any admissible strategy. Then \eqref{def:V} gives us for $K\in[0,T)$
\begin{align*}
\E[U(X_K)] &\le \E[V(yZ_K^\mu) + yX_KZ^\mu_K] \\&
\le \E[V(yZ_K^\mu)] + yx\\
&= \E[V(yZ_K^\mu)] + y\E[Z_K^\mu (yZ_K^\mu)^{1/(p-1)}] = \E[U(X^{(K)}_K)],
\end{align*}
where the last equality follows from $X^{(K)}_K := I(yZ_K^\mu)$ and
$$
V(b) := \inf_{a\ge0} \Big\{ U(a) -ab\Big\} = U(I(b)) - bI(b),\quad b>0.
$$ 
Finally, we note that all expectations appearing in this proof are finite by Lemma \ref{lem:KO} ensured by $\kappa>0$ being small enough.

$\endproof$

\proof[Proof of Lemma \ref{lem:ex}] We first verify that $Z^\lambda$ satisfies Assumption \ref{ass1}. Since $C_1>0$ and $v_t$ is non-central $\chi^2$-distributed, Novikov's condition is satisfied locally in the sense that we can find $\Delta>0$ such that for $n\in\N$ we have
$$
\E\left[\exp\left(\frac12 \int_{n\Delta}^{(n+1)\Delta} \lambda^2_udu\right)\right] \le C\E\left[\exp\left(\int_{n\Delta}^{(n+1)\Delta} C_2(C_3+v_u)du\right)\right] <\infty,
$$
where $C>0$ is some constant. Minor modifications of the iterative argument in the proof of Lemma \ref{lem:KO} shows the global martingale property of $Z^\lambda$. This also verifies the condition of Lemma \ref{lem:novikov}.

$\endproof$


\end{document}